# Tunable Oscillations in the Purkinje Neuron

Ze'ev R. Abrams[1,2,3†], Ajithkumar Warrier[2†], Yuan Wang[2,3], Dirk Trauner[4]  Xiang Zhang[1,2,3*]

1. Applied Science & Technology, University of California, Berkeley,

2. NSF Nanoscale Science and Engineering Center, 3112 Etcheverry Hall, University of California, Berkeley.

3. Material Sciences Division, Lawrence Berkeley National Laboratory, Berkeley

4. Department of Chemistry and Biochemistry, Ludwig-Maximilians-Universität München (LMU), Butenandtstr. 5–13, 81377 Munich, Germany.

[†] *These authors contributed equally to this work.*

*\*- corresponding author: Xiang@berkeley.edu*

**Corresponding author:**

Correspondence to: xiang@berkeley.edu , Professor Xiang Zhang, NSF Nanoscale Science and Engineering Center, 3112 Etcheverry Hall, University of California, Berkeley 94720-1740, Phone: (510) 642-0390, Fax: (510)-643-2311

ABSTRACT

In this paper, we study the dynamics of slow oscillations in Purkinje neurons *in vitro*, and derive a strong association with a forced parametric oscillator model. We demonstrate the precise rhythmicity of the oscillations in Purkinje neurons, as well as a dynamic tunability of this oscillation using a photo-switchable compound. We show that this slow oscillation can be induced in *every* Purkinje neuron, having periods ranging between 10-25 seconds. Starting from a Hodgkin-Huxley model, we also demonstrate that this oscillation can be externally modulated, and that the neurons will return to their intrinsic firing frequency after the forced oscillation is concluded. These results signify an additional functional role of tunable oscillations within the cerebellum, as well as a dynamic control of a time scale in the brain in the range of seconds.

I. INTRODUCTION

The Purkinje Neuron (PN) is the largest neuron in the cerebellum, with over 100,000 inputs and a single output axon [1,2]. Due to its geometry and orientation in the cerebellum, it has been cited as a possible integrator for the motor control system of the brain [2], with many basic neuroscience and artificial intelligence theories based on its complex neuronal network [3,4]. While most studies of the PN focus on biological sources of memory (plasticity) [5,6], a number of studies also describe the functionality of the cerebellum in terms of independent oscillators [6-8].

We have previously reinforced this set of theories with an experimental study demonstrating the intrinsic firing characteristics of the PN [9,10]. We identified three frequency bands inherent to the PN, which we denoted as the Sodium ($Na^+$; >30 Hz), Calcium ($Ca^{2+}$; 1-10 Hz) and Switching bands (>1 Hz). This set of frequency bands is distinct from other regions of the brain [11-14], with the 'Switching' frequency described and measured for the first time [9]. This Switching frequency operates at lower frequencies than those typically associated with memory and other cerebellar processes [14], however there have been recent *in vivo* experiments that have demonstrated similar slow oscillations between 0.039-0.078 Hz [15]. Slow oscillations of firing and quiescence can be defined as an *astable* mode, as opposed to the known *bistable* mode in the PN [16-18], and we have recently described a new theory of PN function using the terminology of electronic oscillator systems exhibiting astable, bistable and monostable modalities [10].

In this paper, we first show that every PN can exhibit this slow form of astable oscillation when activated using pharmacological compounds *in vitro*. These slow oscillations are shown to be precise, with high quality factors of resonance. Next, we modulate the frequency of these neurons using a unique form of highly specific, photo-switchable compound [19]. By doing this, we show that this frequency pattern acts as a forced-oscillator when externally driven, and that the oscillations revert back to their initial frequency once the driving force is stopped. Using the Hodgkin-Huxley neuron model [20], we derive a form of parametric oscillator that describes the slow oscillations observed, as well as their ability to be externally tuned. Finally, we analyze the parameters of oscillation, and compare them with the existing literature for further understanding of the gating mechanism in the oscillation behavior of Purkinje neuron cells.

## II. METHODS

For a complete description of animal handling, sample preparation, solutions used and the optical/patch-clamp setup, please refer to [9].

The stimulation of PNs was done by activating kainate receptors using a variety of highly selective molecular agonists (KRAs). These molecules act only upon kainate receptors without activating any of the other glutamatergic receptors on the cell, particularly AMPA (α-amino-3-hydroxyl-5-methyl-4-isoxazole-propionate) receptors, which are the majority of ionotropic glutamatergic receptors on PNs. Additional pharmacological blockers were used to isolate the kainate response in the PN. The Photo-Switchable Kainate Receptor Agonist (PSKRA) [19] was used as a traditional KRA when in the dark. The photo-response is described in [19], as well as in the text.

All drugs except the Photo-Switchable Kainate Receptor Agonist (PSKRA) were purchased via Sigma-Aldrich or Tocris Bioscience. Drugs were applied to the artificial cerebrospinal fluid (ACSF) reservoir and allowed to perfuse onto the slice using the closed-loop system. Kainate activation of the PNs was achieved using either highly-specific kainate receptor agonists (KRAs), or Monosodium Glutamate (MSG, 100 μM) in conjunction with an AMPA receptor blocker GYKI-52466 (10-20 μM). The KRAs used consisted of the commercially available (2S,4R)-4-Methylglutamic Acid (SYM-2081, 10-50 μM), a non-selective GluK1/GluK2 agonist (non-selective for GluR5/6, selective over AMPA receptors), as well as the PSKRA. The PSKRA was based upon a variant of the commercially available SYM-2081, called LY339434 [21], which was designed specifically to be selective towards GluK1 (GluR5) over both GluK2 (GluR6) and AMPA receptors, and was used at 50-100 μM, the ideal concentration as described in [19].

## III. RESULTS

### A. Induction of Astability

We first demonstrate that every PN, when measured *in vitro* using a current-clamp setup, will change its firing pattern to that consisting of slow oscillations of $Ca^{2+}$ spikes [22], nested within a slow Switching envelope wave. Fig. 1(a) displays the recording of a cell transitioning to this mode (after the application of KRAs in conjunction with teotrodotoxin (1 μM), which abolishes the $Na^+$ spikes, and accentuates the underlying $Ca^{2+}$ spike pattern). The transition to a slow oscillation mode consisting of Switching and Calcium frequencies occurs after 0.5-2 min, as shown on the right side of Fig. 1(a). Once induced into this slow oscillation mode, the system retains its astability, with cells oscillating between firing $Ca^{2+}$ spikes and quiescence for up to 40 min. The oscillation frequency can be measured from the highest peak in the Fourier Transform (FFT) of the recording, as shown in Fig. 1(b). Results such as these were obtained in *n*>50 cells, with a high yield of greater than 90%, signifying the reproducibility of these results.

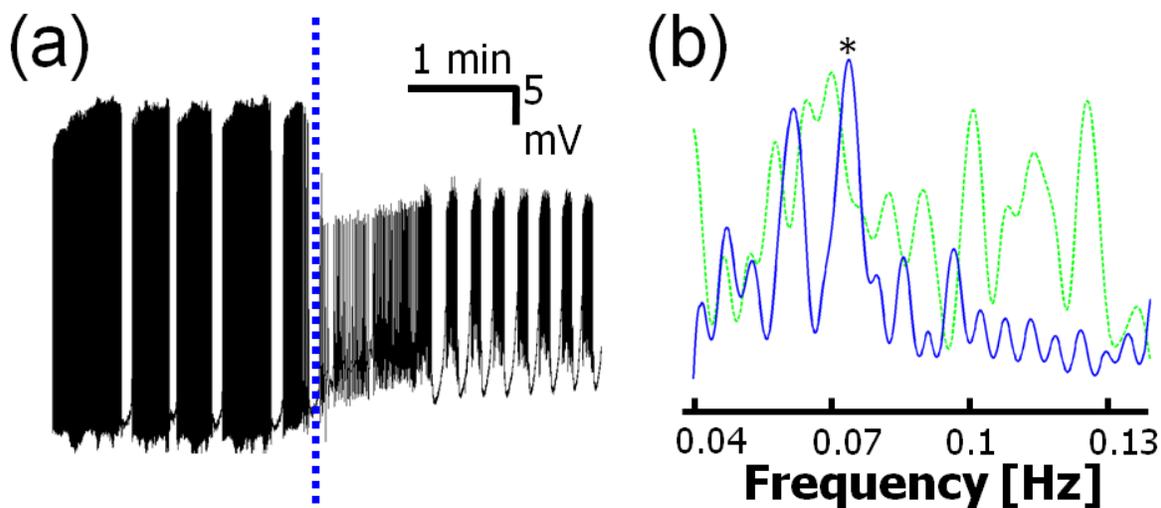

**Figure 1: (Color online) Astability induction in a Purkinje neuron.**

(**a**) Signal output recording of a cell transitioning to an astable oscillatory mode after applying a kainate receptor agonist (PSKRA, 100 µM) in conjunction with an AMPA receptor blocker (GYKI, 10 µM) and $Na^+$ channel blocker (TTX, 1 µM). Dotted line signifies the drugs' arrival at the cell within the bath chamber. Clear oscillations are at the far right. (**b**) Fourier transform of the recording before (dashed green) and after (solid blue) the induction of oscillations. A clear peak at 0.075 Hz can be seen (asterisk).

The frequency of the astable oscillations can be measured by tracking the on/off transitions in time using a windowing algorithm [9], or directly by measuring the Fourier transform using the FFT algorithm (Matlab). This is shown in Fig. 2 for a cell, already in the astable mode. Both window-tracking [Fig. 2(b)] and the FFT [Fig. 2(c)] display the clear astable oscillation frequency of the cells, with the precision of the oscillation measured as the deviation from the mean in the time domain, or as the Full Width at Half Maximum (FWHM) in the frequency domain. The transition to clear, precise oscillations would typically occur within 0.5-2 min of the drugs' activation upon the cell, as is seen in the transition region in Fig. 1(a), as well as the initial firing sequence in Fig. 2(a).

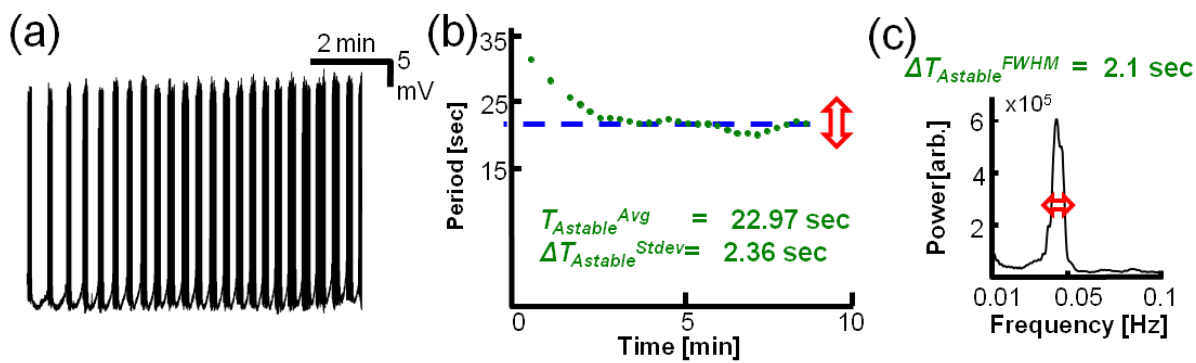

**Figure 2: (Color online) Measuring the induced oscillations.**

(**a**) 9 minute recordings of a cell induced to oscillate via application of glutamate (MSG, 100 µM) and an AMPA receptor blocker (GYKI, 10 µM). The period of oscillation of bursts can be tracked over time (**b**), or examining the low frequency region of the frequency domain (FFT) of the recording (**c**). The precision of the oscillation frequency can be measured from the center of the peak in (**c**), with the Full Width at Half Maximum (FWHM) signifying the variation over time, or as the standard deviation from the mean period.

## B. Tunability via Photo-Switching

Until now we have shown only the natural astability frequencies measured in PNs after pharmacological activation; presently we show that the PN can act as a *forced* oscillator, and thus extend the frequency range available for astability. This can allow the PNs to produce a wider range of frequencies per cell, which can then be integrated together in other regions of the cerebellum [10,23]. Using the PSKRA and the appropriate photo-switching wavelengths, shown in Fig. 3(a), we can gradually toggle the firing of $Ca^{2+}$ spike bursts in a PN to a given oscillation period [Fig. 3(b)]. We observed that the firing pattern then matches the period of forced-modulation. This was done at a wide range of periods (6-30 sec) and DCs (25-75%).

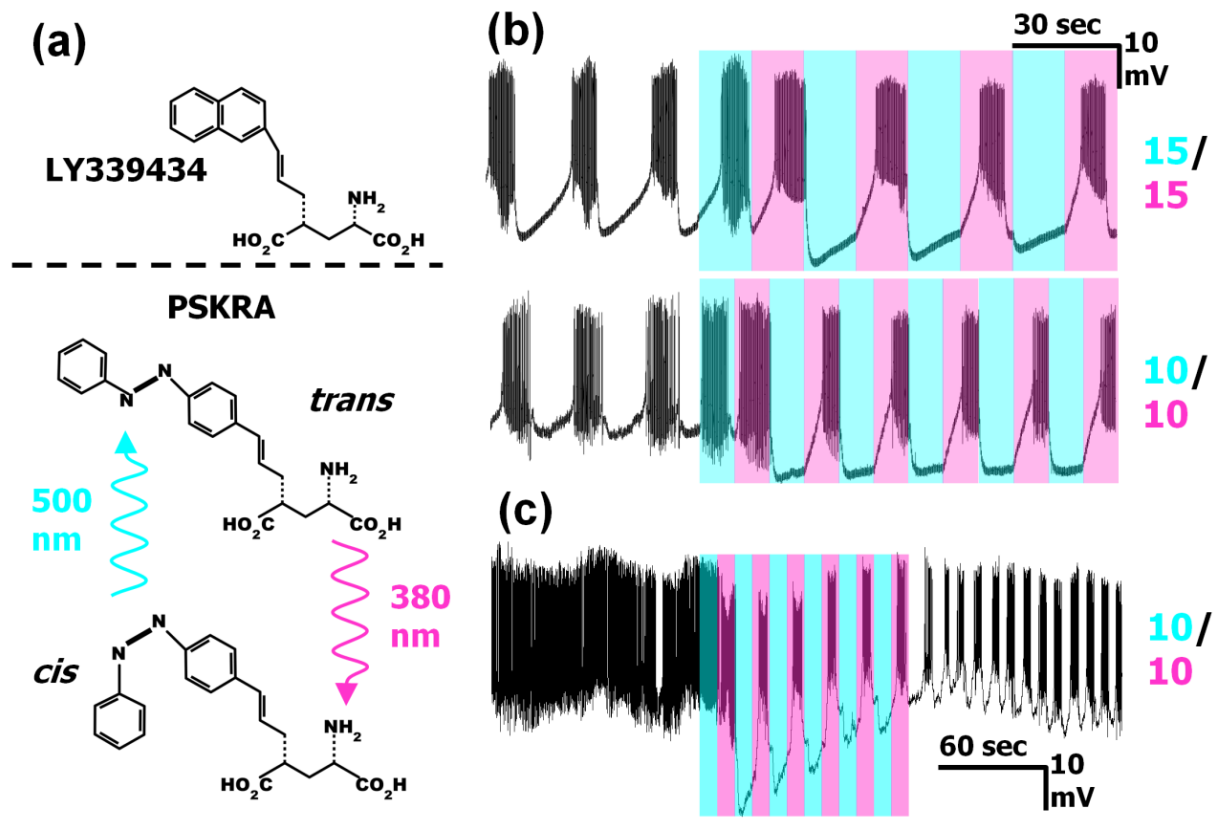

**Figure 3. (Color online) Photo-switchable compound modulates the oscillation.**

(**a**) Comparison of the GluK1 selective compound LY339434 and the PSKRA, which has an added azobenzene moiety, rendering it photo-switchable at 500 and 380 nm. (**b**) The ultraviolet (UV, 380 nm) and cyan (500 nm) light turn the bursting on and off, respectively, as a function of the modulation period, here shown for two modulation patterns with differing periods. The stimulation pattern is noted to the right of each recording, with stimulations at 15 sec cyan, 15 sec UV and 10 sec cyan, 10 sec UV. (**c**) Photo-modulation followed by induction of oscillation in a cell not displaying astability prior to the modulation. The cell continued to oscillate after the photo-switching was completed.

The tunability of cells using the PSKRA was effective in all cells measured, with the slow oscillations following the forced photo-modulation. Additionally, in some of the cells that did not exhibit astability initially when in the presence of the PSKRA in the dark ($n=5$, resulting in the less than 100% yield reported above), the photo-modulation still resulted in forced-oscillation, with the cells continuing to oscillate after the photo-switching was stopped, as demonstrated in Fig. 3(c).

### C. Delayed Recovery of Modulation

The dynamics of the recordings subsequent to the photo-modulation were also studied. Figure 4(a) displays a representative recording from a PN that was initially oscillating at a natural period of 22 sec (green), and then photo-modulated for 2 min at 5 sec on/ 5 sec off (10 sec period, dark blue), and subsequently allowed to recover (light blue). The PN faithfully follows the forced-modulation after 1-2 stimuli when the light is applied, and then slowly recovers back to the natural oscillation period it had prior to the stimulation, after the photo-stimulation is stopped. This recovery response of the cell's oscillatory modulation is similar to a traditional forced-oscillator, with an exponential recovery after the forced modulation is stopped. This is best visualized when plotting out the period and DC over time, as in Figs. 4(b) and (c), which display both the period-matching during forced-oscillation (dashed blue line) as well as the exponential recovery (dotted red line).

The time constants for recovery have a distribution among different cells, with a recovery time of $82 \pm 48$ sec for the period, and $70 \pm 41$ sec for the DC ($n=11$ cells. Errors are in SD, displaying the range of variation among cells). The long recovery time for this process is similar to other forms of short-term memory in the brain (such as short-term depression and potentiation [5]), allowing the cell to "remember" its forced-modulation for a short duration after the stimulus is applied, however the cells typically did not retain the forced frequency. The direction of recovery was generally towards the natural astable period in the cell prior to photo-modulation, with forced-photo-modulation of the cells done both below and above their natural period [below: 5/5 sec on/off modulation, $n=7$; above: 10/10 sec on/off modulation, $n=9$; 15/15 sec on/off modulation, $n=4$, with representative cell recordings in Figs. 4(d) and (e), respectively].

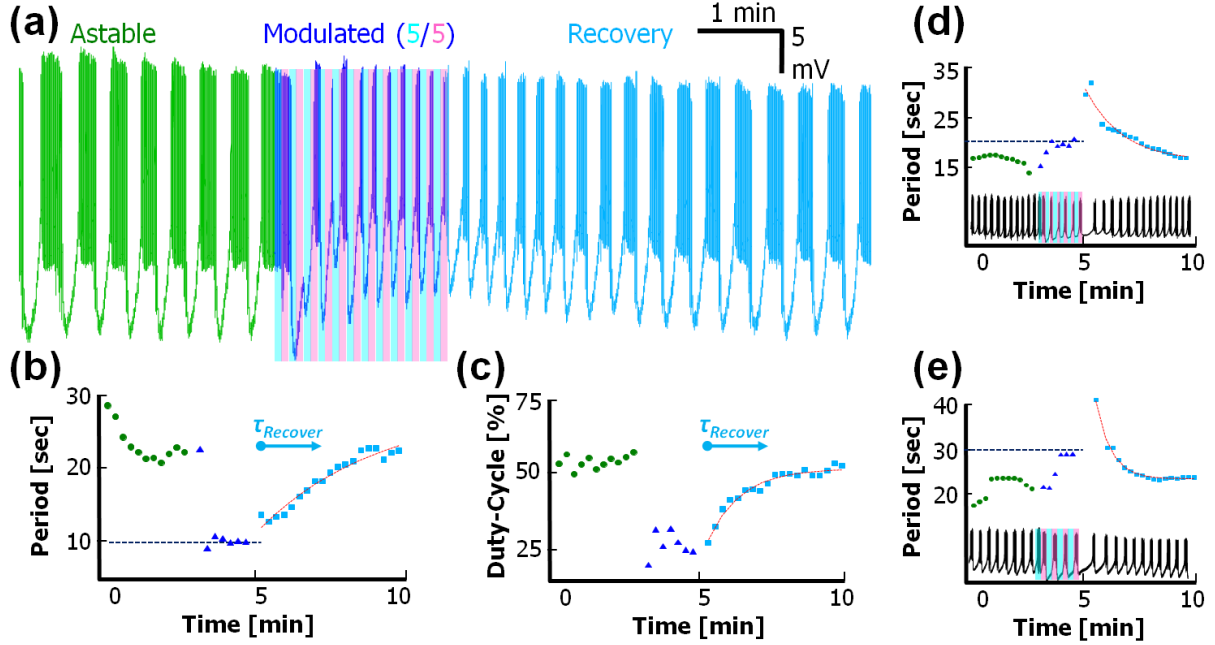

**Figure 4: (Color online) Recovery from the photo-modulation follows a forced-oscillator model.**

(**a**) Exemplary recording of a Purkinje neuron showing a natural oscillation (green) that is then photo-modulated at 5 sec on/5 sec off (10 sec period) for 2 min (dark blue), and then allowed to recover from the modulation (light blue). (**b**, **c**) Period and Duty-Cycle tracking of the cell in (**a**), displaying the forced-modulation (dashed blue line) and exponential recovery. Dots are color coded as in (**a**). Red dotted lines are fitted exponential curves for the recovery segment, with a time constant of $\tau = 124$ sec and $\tau = 67$ sec for the period and duty-cycle, respectively. (**d**, **e**) Representative plotted periods of the natural, modulated and recovering oscillators for two different modulations: (**d**), 10/10; (**e**), 15/15 on/off, in seconds ($\tau = 97$, 40 sec, respectively). Actual recordings appear below each plot.

## IV. PARAMETRIC OSCILLATOR MODEL

The bursting oscillations of the PNs are non-dissipating, with a clear frequency signature in the FFTs. We can simplify the Hodgkin-Huxley equations [24-27] to reach a model similar to the fundamental parametric oscillator equations to describe this harmonic signal. The generalized formula for the membrane potential, *V*, injected current, *I*, and membrane capacitance, *C*, is [25,26]:

$$C\dot{V} = -I - \sum g_i x_i^k (V - E_i) \qquad (1)$$

With each gating variable, $x_i$, being a function of the voltage: $\dot{x} = [x_0(v) - x]/\tau(v)$, the parameters, *g*, being the conductance, and ($V$-$E_i$) term being the driving force per ion channel. Taking a time derivative of equation 1:

$$C\ddot{V} = -\dot{I} - \sum \left[ \dot{g}_i x_i^{ki} (V - E_i) + g_i \frac{d(x_i^{ki})}{dt}(V - E_i) + g_i x_i \dot{V} \right] \qquad (2)$$

Assuming that the injected current (*I*, if any exists) is constant with time, and that the conductance coefficients $g_i$, are also time independent, we can neglect the *dI/dt* and $dg_i/dt$ terms, remaining with:

$$C\ddot{V} = -\sum\left[ g_i \frac{d(x_i^{ki})}{dt}(V-E_i) + g_i x_i \dot{V} \right] \quad (3)$$

Since we are interested in the slow changing terms only, whose changes with time are of the order of the oscillation period (10-25 sec), we can isolate these terms by removing them from the summation:

$$C\ddot{V} = g_{Slow}\frac{d(x_{Slow}^S)}{dt}(V-E_{Slow}) + g_{Slow} x_{Slow}^S \dot{V} + \sum_{fast}\left[ g_i \frac{d(x_i^{ki})}{dt}(V-E_i) + g_i x_i \dot{V} \right] \quad (4)$$

In equation 4, the *Slow* gating variable has an exponent of *S*. We will assume that this exponent is unity, since the slow action of our experimental results is similar to that of either the muscarinic gating variable *M*, or *h* for the $I_h$ refractory current, both which have an exponent of unity [25, 26]. This assumption will be further justified in the subsequent section.

The rightmost bracketed term in the summation can be taken as the *average value* for long periods where the fast acting terms within these brackets change at a rate that is much higher than the slow oscillations examined here. This is because the time derivative *d/dt* for long time durations is defined by changes of the order of $1/\Delta t_{Slow}$. This is comparable with taking the average membrane potential during the firing of action potentials, otherwise known as the "up" state of a bistable system [16], such that we are dealing only with a *slow* waveform that is similar to a square-wave envelope. With these assumptions, we obtain:

$$C\ddot{V} = g_{Slow}\dot{x}_{Slow}V - g_{Slow}\dot{x}_{Slow}E_{Slow} + g_{Slow}x_{Slow}\dot{V} + \left(\sum_{fast}\left[ g_i \frac{d(x_i^{ki})}{dt}(V-E_i) + g_i x_i \dot{V} \right]\right)_{avg} \quad (5)$$

The middle term: $g_{Slow}x'_{Slow}E_{Slow}$ can further be isolated, since it is a term that is dependent upon the voltage and time, but does not directly include the voltage. Reorganizing the above equation, we can obtain a generalized 2$^{nd}$ order differential equation:

$$C\ddot{V} - g_{Slow}x_{Slow}\dot{V} - g_{Slow}\dot{x}_{Slow}V = -g_{Slow}\dot{x}_{Slow}E_{Slow} + \left(\sum_{fast}\left[ g_i \frac{d(x_i^{ki})}{dt}(V-E_i) + g_i x_i \dot{V} \right]\right)_{avg}$$
$$(6)$$

Or:

$$C\ddot{V} - g_{Slow}x_{Slow}\dot{V} - g_{Slow}\dot{x}_{Slow}V \approx F(\bar{t}) \quad (7)$$

Comparing this to the harmonic oscillator with a driven source, *F(t)* [28]:

$$m\ddot{V} - b\dot{V} + kV = F(t) \quad (8)$$

Equation 8 is the classic equation for a *driven harmonic* oscillator with a resonant frequency of $\omega_o^2=k/m$ and a quality factor of $Q=\omega_o m/b$, with $F(t)$ being the time-dependent driving input. Comparing terms, we find that:

$$\begin{aligned} m &= C \\ b(V,t) &= g_{Slow} x_{Slow} \\ k(V,t) &= g_{Slow} \dot{x}_{Slow} \end{aligned} \quad (9)$$

As can be seen, the parameters of this equation are *time/voltage dependent*. This makes the equation a *parametric* oscillator as opposed to a simple *harmonic* oscillator. The parameters in Equation 9 are related to the biological and measureable aspects of each neuron, with the membrane capacitance directly measureable, and the gating variables measureable using voltage-clamp experiments to determine the dynamics of the ion channels involved.

If the oscillator is under-damped ($Q>1/2$, which is equivalent to $b<<\omega_o$), it is easy to then measure the resonance frequency of the Purkinje neuron ($2\pi f_o=\omega_o$) as well as the quality factor, $Q$, from the Signal-to-Noise-Ratio (SNR). The frequency of an oscillating Purkinje neuron can be measured directly in the frequency domain, and the SNR can be calculated either through the time-tracking algorithm, or from the width of the peak in the FFT. The frequency of the damped oscillator is quite similar to that of the resonance frequency: $\omega^2_{measured}=\omega_o^2(1-1/4Q^2)$, with $Q$ taken as the SNR. Since most of the oscillating neurons had $Q>2$ (as will be shown subsequently), then $\omega_{measured}\approx\omega_o$. Therefore, by measuring these two parameters ($f_o$ and $Q$), and taking a general value of the membrane capacitance of $C=1$ µF/cm$^2$ [26], we can obtain the parameters $k$ and $b$.

Equation 7 is the parametric oscillator equation for an oscillating neuron that can be analyzed using phase-space diagrams [27]. Obtaining the van-der-Pol oscillator equation from the Hodgkin-Huxley equations follows a similar method.

## V. DATA ANALYSIS

The period (1/frequency) and duty-cycle (DC) of the astable mode were measured in $n=43$ cells, each oscillating for at least 7 min, and are displayed in Figs. 5(a) and (b). Since each cell acts as an independent oscillator, it is expected to find a range of inherent frequencies. We found that the average period of the cells was $20 \pm 8$ sec, and the average DC was $46 \pm 8\%$ ($\pm$ standard deviation, SD). We again note the similarity between the range of slow oscillations measured here, and those measured *in vivo* in tottering mice, which ranged between 12.82 and 25.64 sec [15].

The precision of oscillation over time can be measured by the quality factor of the resonator, $Q$, or the SNR. This was measured using the window-tracking algorithm in the time domain, for each of the cells measured, and displayed in Fig. 5(c) for each individual cell. Of the PSKRA activated cells, 45% ($n=13/29$, black circles) had a SNR larger than 10, corresponding to less than a 10% deviation in period over time, whereas the SYM-2081/MSG activated cells were less accurate (blue triangles). This may be attributable to wither the selectiveness of the PSKRA, or the concentration ratios.

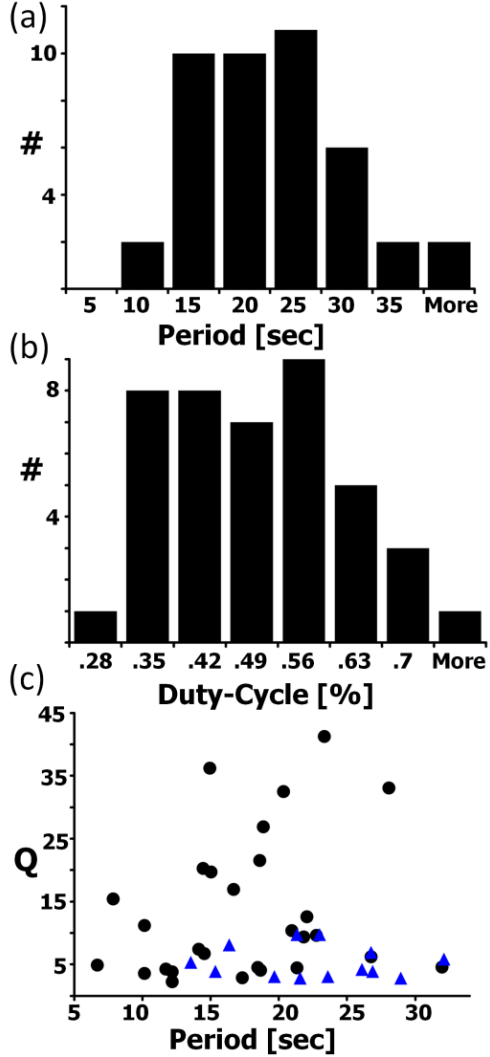

**Figure 5: (Color online) Period, duty-cycle and precision of oscillation.**

(**a**, **b**) Histograms of the average period and duty-cycle (*n=43* cells total) showing the central frequency of 20 sec/0.05 Hz, and 46% duty-cycle. (**c**) Quality factor (*Q*), or SNR, data for the PSKRA (black circles) and SYM-2081/MSG (blue triangles) activated cells, as a measure of the precision of oscillation. Each cell is plotted individually, with the PSKRA activated cells having higher precision values.

Using the formalism of the parametric oscillator above, we can view each PN as an independent oscillator, each with its own measurable parameters of oscillation. For example, using the cell shown in Fig. 2, we can isolate the frequency and quality factor directly from the peak in the FFT in Fig. 2(c): with $Q=10.93$ and $f_o=0.044$ Hz ($T=22.9$ sec); these result in values of $b=0.025$ and $k=0.07$. Following equation 9, since we are measuring time averaged values, we can assume that the gating variables vary sinusoidally (using Floquet analysis), such that the time averaged value is half the amplitude: $<b>=g_{Slow}\times<x_{Slow}>=g_{Slow}/2$. This results in a value of $g_{Slow}\sim 0.05$ mS/cm$^2$. Comparing with existing values in the literature would place this gating channel closer to the h-current with $g_{Ih}=0.03$ mS/cm$^2$ [25] or perhaps the muscarinic channel with $g_M=0.75$ mS/cm$^2$ [26]. The relationship between the h-current and bistability has previously been shown [16].

## VI. CONCLUSIONS

This work has shown the capability of a PN to act as an astable oscillator with long periods of oscillation (10-25 sec), as well as the ability to externally tune this frequency for extended periods of time. This frequency range is notably outside the range typically studied in the brain [11-14], but matches other *in vivo* results of the PN [15]. The assumed existence of such a timing functionality of the PN in the cerebellum lies in complete agreement with the conceptual view of the cerebellum as the feedback control mechanism of the brain [1,6], as well as temporal pattern generator theories of the cerebellum [23]. Using mathematical dynamic systems models and newly derived optical activation techniques allow us to probe the intrinsic behavior of cells within a network, thereby enabling us to reverse engineer the neuronal circuitry of the brain at a higher level.

## ACKNOWLEDGMENTS

This research was made with support from the National Science Foundation Nano-Scale Science and Engineering Center (NSF-NSEC) under award CMMI-0751621. ZRA acknowledges Government support under and awarded by DoD, Air Force Office of Scientific Research, National Defense Science and Engineering Graduate (NDSEG) Fellowship, 32 CFR 168a. ZRA would also like to thank Prof. Harold Lecar for his useful discussion.